\documentclass{jfm}
\usepackage{color}
\usepackage{amstext}
\usepackage{amssymb}
\usepackage{graphicx}
\usepackage{amsmath}

\newcommand\const{\mathrm{const}}

\newcommand\vC{\boldsymbol{C}}
\newcommand\vX{\boldsymbol{X}}

\newcommand\vF{\boldsymbol{F}}

\newcommand\vV{\boldsymbol{V}}

\newcommand\va{\boldsymbol{a}}
\newcommand\vb{\boldsymbol{b}}

\newcommand\vf{\boldsymbol{f}}

\newcommand\vl{\boldsymbol{l}}

\newcommand\vn{\boldsymbol{n}}

\newcommand\vx{\boldsymbol{x}}

\newcommand\vq{\boldsymbol{q}}
\newcommand\vQ{\boldsymbol{Q}}
\newcommand\vg{\boldsymbol{g}}

\newcommand\vxi{\boldsymbol{\xi}}

\newcommand\vsigma{\boldsymbol{\sigma}}

\begin{document}

{\title[Dumbbell micro-robot driven by flow oscillations] {Dumbbell micro-robot driven \\by flow oscillations
}}

\author[V. A. Vladimirov]
{V.\ns A.\ns V\ls l\ls a\ls d\ls i\ls m\ls i\ls r\ls o\ls v}

\affiliation{Dept of Mathematics, University of York, Heslington, York, YO10 5DD, UK}

\pubyear{2012} \volume{xx} \pagerange{xx-xx}
\date{Oct 21st 2012}

\setcounter{page}{1}\maketitle \thispagestyle{empty}

\begin{abstract}

In this paper we study the self-propulsion of a dumbbell micro-robot submerged in a viscous fluid. The
micro-robot consists of two rigid spherical beads connected by a rod or a spring; the rod's/spring's length
is changing periodically. The constant density of each sphere differs from the density of a fluid, while the
whole micro-robot has neutral buoyancy. An effective oscillating gravity field is created \emph{via}
rigid-body oscillations of the fluid. Our calculations show that the micro-robot undertakes  both
translational and rotational motion. Using an asymptotic procedure containing a two-timing method and a
distinguished limit, we obtain analytic expressions for the averaged self-propulsion velocity and  averaged
angular velocity. The important special case of zero angular velocity represents rectilinear self-propulsion
with constant velocity.

\end{abstract}

\section{Introduction \label{sect01}}

The study of self-propelling micro-robots is a flourishing modern research topic,  striving to create a
fundamental base for modern applications in medicine and technology, see \emph{e.g.}
\cite{Purcell, Koelher, NG+, Dreyfus, Felderhof, Yeomans1, Paunov, Lefebvre,  Gilbert, Golestanian, Yeomans,
Pietro, Belovs, Lauga, Romanczuk}. We define \emph{self-propulsion} as the motion of a micro-robot which is
subjected to zero external total force. The simplicity of micro-robots' geometry represents a major advantage
in contrast to the extreme complexity of self-swimming microorganisms. This advantage allows us to describe
the motions of micro-robots in  greater depth.
\begin{figure}
\centering\includegraphics[scale=0.7]{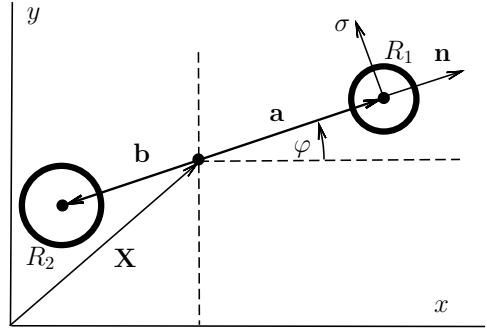}
\caption{Two spheres, linked by a rod of periodically changing length.}
\label{D-robot}
\end{figure}
The major problem in the designing  of a micro-robot is the need for an external source of energy to provide
for its oscillatory behaviour. Proposed sources include an oscillating (or rotating) magnetic field (see
\cite{Dreyfus, Belovs, Gilbert}), an electric field (see
\cite{Paunov}), and even molecular Brownian forces (see  \cite{Romanczuk}). At the same time, the major
oscillatory forces available in  fluid have not been exploited; these are the forces caused by fluid
oscillations which are imposed by periodically varying boundary conditions, waves, or turbulence. The ratio
of characteristic spatial scales (several microns for a micro-robot \emph{vs.} millimeters, centimeters, or
greater scales for flow oscillations) makes it clear that the first problem to study is the behaviour of a
micro-robot in a fluid that oscillates  as a rigid body.

In this paper, we consider the self-propulsion of a two-sphere \emph{Buoyancy}-driven \emph{Dumbbell}
micro-robot (we call it \emph{BD}-robot, see the figure). The whole micro-robot is neutrally buoyant (in
order to avoid sedimentation);  one of its spherical beads is positively buoyant and the other is negatively
buoyant. We study two versions of \emph{BD}-robots. In the first one the beads are connected by a rod of
prescribed oscillating length, in the second one the beads are linked by an elastic spring. First, we study
the case of a rod and, next,  we consider the changes that appear after replacing  the rod with a spring. A
mathematical formulation of the problem leads us to the study of creeping motions with time-periodic forces.
The problem is solved by employing a version of the two-timing method and distinguished limit arguments,
developed in
\cite{Vladimirov0,Vladimirov1,VladimirovMHD}. The approach allows any motion of \emph{BD}-robot to be described
analytically. Our calculations show that, generally, the \emph{BD}-robot participates in both translational
and rotational motion.  Rectilinear translational self-propulsion with constant velocity represents a special
case of this solution. We have calculated the velocity of rectilinear self-propulsion and the ranges of
governing parameters that correspond to translational motions.

\section{Problem formulation \label{sect01}}

The \emph{BD}-robot represents a dumbbell configuration, which consists of two homogeneous rigid spherical
beads of different radii $R_\nu$, $\nu=1,2$ connected by a rod of length $l$ (see the figure).   We study
two-dimensional motions of a tree-dimensional dumbbell in cartesian coordinates $(x,y)$. The centers of the
spheres $\vx^{(\nu)}$ are described as
\begin{eqnarray}
&&\vx^{(1)}=\vX+\va,\quad \vx^{(2)}=\vX+\vb,\quad R_1\va+R_2\vb=0 \label{notations}\\
&&\va=a\vn=r_1 l\vn,\ \vb=b\vn=-r_2 l\vn;\quad r_1\equiv R_2/(R_1+R_2),\ r_2\equiv R_1/(R_1+R_2)\nonumber
\end{eqnarray}
where $\vX=(X,Y)$ is the radius-vector of a center of reaction. The axis of symmetry of a dumbbell is given
by the vector  $\vl\equiv\vx^{(1)}-\vx^{(2)}$ , $l\equiv|\vl|$. The unit vectors ${\vn}$, ${\vsigma}$ and the
angle $\varphi$ are given by
\begin{eqnarray}
\label{units}
&&{{\vn}}\equiv
\left(
\begin{array}{c}
    \cos\varphi  \\
     \sin\varphi
  \end{array}
\right),\quad
{{\vsigma}}\equiv
\left(
\begin{array}{c}
    -\sin\varphi  \\
     \cos\varphi
  \end{array}
\right),\quad \vsigma=\vn_\varphi,\quad \vn=-\vsigma_\varphi,\quad \vn\cdot\vsigma=0
\end{eqnarray}
where the subscript $\varphi$ stands for $d/d\varphi$. The length $l$ is changing periodically
\begin{eqnarray}
&& l= L+\varepsilon\widetilde{l}(\tau);\
\tau\equiv\omega t;\quad \omega=\const, \ \varepsilon=\const
\label{constraint}
\end{eqnarray}
where $L$ is a constant averaged value and $\widetilde{l}$ is a $2\pi$-periodic function of $\tau$ with zero
average value (throughout the paper a `tilde' above a function of time denotes that this function is
oscillating and has zero mean value). The spheres experience external friction forces
$\vF^{(\nu)}=(F_1^{(\nu)},F_2^{(\nu)})$ while the rod is so thin (in comparison with either  $R_\nu$) that
its interaction with the fluid can be considered negligible.

We consider the motions of a \emph{BD}-robot in a viscous incompressible fluid which, in the absence of the
\emph{BD}-robot, oscillates as a rigid body. These rigid-body oscillations are prescribed as a two-dimensional
translational spatial displacement $\widetilde{\vxi}(\tau)=(\xi_1(\tau),\xi_2(\tau))$ of fluid particles (at
infinity in space); the related acceleration is $\widetilde{\vxi}_{tt}=\omega^2\widetilde{\vxi}_{\tau\tau}$,
where the subscripts stand for related derivatives. The problem can be studied in an oscillating
(non-inertial) system of reference, in which a fluid at infinity is in a state of rest. In this frame,
according to Einstein's principle of equivalence, or according to a related transformation of a Lagrangian
function, the equations of fluid motion are standard, however, they contain an additional oscillating gravity
force
\begin{eqnarray}\label{gravity}
\widetilde{\vg}=-\omega^2\widetilde{\vxi}_{\tau\tau}
\end{eqnarray}
which causes buoyancy forces  $\widetilde{\vf}_b^{(\nu)}=-M^{(\nu)}\widetilde{\vg}$, where the coefficient
$M^{(\nu)}$ is equal to the difference in the mass of a sphere and the mass of a displaced fluid; $M^{(\nu)}$
can be either positive or negative. The potential energy of a sphere is
$\Pi^{(\nu)}=M^{(\nu)}\widetilde{\vg}\cdot\vx^{(\nu)}$. We consider a \emph{BD}-robot of neutral total
buoyancy, with total potential energy
\begin{eqnarray}\label{Pi}
\Pi=\Pi^{(1)}+\Pi^{(2)}=M^{(1)}\widetilde{\vg}\cdot\vl,\quad M\equiv M^{(1)}=-M^{(2)}>0
\end{eqnarray}
The problem formulation contains three characteristic lengths: the length of the rod $L$, the radius of the
spheres $R$, and the amplitude  of the rod's oscillation $a$. In addition we have the characteristic
time-scale $T$, excess mass $M$,  gravity  $G$, and  viscous force $F$. We have chosen these scales as
$R\equiv(R_1+R_2)/2$, $T\equiv 1/\omega$, $a\equiv\varepsilon L$, $F\equiv 6\pi\eta RL/T$,
$g\equiv\max|\widetilde{\vg}(\tau)|$, where $\eta$ is the fluid viscosity. The dimensionless  variables
(marked with asterisks) are $\vx=L\vx^*$, $t=T\,t^*$, $F_i=F\,F_i^*$. Three independent small parameters of
the problem are
\begin{eqnarray}\label{small-par}
&&\varepsilon\equiv a/L,\quad \delta\equiv 3R/(4L),\quad m\equiv Mg/F
\end{eqnarray}
Below we use only dimensionless variables, however all asterisks are omitted. Note, that in the chosen
dimensionless units, $R_1+R_2=2$.

We choose the generalised coordinates of \emph{BD}-robot to be $\vq=(q_1,q_2,q_3,q_4)\equiv(X,Y,l,\varphi)$.
The motion of the \emph{BD}-robot, with a given $l(\tau)$ (\ref{constraint}), is described by the Lagrangian
function $\mathcal{L}=\mathcal{L}(\vq,\vq_t)$, which includes the constraint (\ref{constraint}) with
lagrangian multiplier $\lambda$
\begin{eqnarray}
&&\mathcal{L}(\vq,\vq_t)=\mathcal{K}-\Pi+\lambda(l-L-\varepsilon \widetilde{l}\,)
\label{Lagr-contstr}
\end{eqnarray}
where $\mathcal{K}$, and $\Pi$ are kinetic energy and potential energy (\ref{Pi}) of  \emph{BD}-robot;
$\lambda$ represents an additional unknown function of time. The Lagrange equations are
\begin{eqnarray}
&&\frac{d}{dt}\frac{\partial \mathcal{L}}{\partial q_{n t}} -\frac{\partial \mathcal{L}}{\partial
q_n}=Q_n,\quad Q_n=\sum_{\nu=1}^2\sum_{k=1}^2 F_k^{(\nu)}\frac{\partial x_k^{(\nu)}}{\partial
q_n}\label{Lagr-eqns}
\end{eqnarray}
where $\vQ=(Q_1,Q_2,Q_3,Q_4)$ is the generalized external viscous force, exerted by a fluid on the
\emph{BD}-robot. As one can see, we use latin subscripts ($i,k=1,2$) for cartesian components of vectors and
tensors, subscript $n=1,2,3,4$ for generalised coordinates, and subscripts (or superscripts) $\mu,\nu=1,2$ to
identify the spheres.

We accept, that the fluid flow past \emph{BD}-robot is described by the Stokes equations, where all inertial
terms are neglected. In the consistent approximation the mass of a rod and masses of the spheres are
negligible, hence $\mathcal{K}\equiv 0$. Therefore (\ref{Lagr-contstr}),(\ref{Lagr-eqns}),(\ref{notations})
give use to the following system of equations
\begin{eqnarray}
&&\vF^{(1)}+\vF^{(2)}=0\label{sum-forces}\\
&&\varepsilon\alpha\vg\cdot\vsigma=\vF^-\cdot \vsigma,\quad 2\vF^-\equiv R_2\vF^{(1)}-R_1\vF^{(2)}\label{sum-torques}\\
&&\varepsilon\alpha\vg\cdot\vn-\lambda=\vF^-\cdot \vn
\label{eqns-reactions}
\end{eqnarray}
which is supplemented by  constraint (\ref{constraint}). The great advantage of a Lagrangian formalism is its
self-sufficiency. In particular, the conditions of zero force (\ref{sum-forces}) and the balance of torques
(\ref{sum-torques}) appear automatically, while eqn.(\ref{eqns-reactions}) allows us to find the reaction of
constraint $\lambda$. In (\ref{sum-torques}),(\ref{eqns-reactions}) we have accepted that
$m=\varepsilon\alpha$ with  constant $\alpha=O(1)$; this is our physical assumption, which states that two
small parameters $\varepsilon$ and $m$ (out of three parameters in the list (\ref{small-par}))  are of the
same order. Physically, it means that the difference between the densities (of each sphere and the fluid) or
the amplitude of oscillations of the fluid is small (or both these parameters are small). The explicit
expressions for $\vF^{(\nu)}$ are
\begin{eqnarray}\label{forces-Stokes}
&&\vF^{(1)}\simeq -R_1\vx_t^{(1)}+\delta \kappa \mathbb{S} \vx_t^{(2)},\quad
\vF^{(2)}\simeq -R_2\vx_t^{(2)}+\delta \kappa \mathbb{S} \vx_t^{(1)}\\
&& l^3\mathbb{S}=l^3 S_{ik}\equiv l^2\delta_{ik}+l_i l_k,\quad \kappa\equiv R_1R_2\nonumber
\end{eqnarray}
Each force $\vF^{(\nu)}$ represents the first approximation (with the error $O(\delta^3)$) for the Stokes
friction force exerted on a sphere moving in a flow field generated by another sphere. To construct
(\ref{forces-Stokes}) we use a classical explicit formula for the fluid velocity past a moving sphere, see
\cite{Lamb,Landau,Moffatt}. Eqns. (\ref{sum-forces})-(\ref{forces-Stokes}) represent a system of four
equations for four unknown functions of time:\ $X$, $Y$, $\varphi$, and $\lambda$.  For the prescribed $l$
(\ref{constraint}), the equation (\ref{eqns-reactions}) need not to be considered if we are interested only
in the motion of the micro-robot  and are not calculating of  reaction force $\lambda$. For future use, we
rewrite (\ref{sum-forces}),(\ref{sum-torques}) as
\begin{eqnarray}
&&\vX_t-\delta \kappa\mathbb{S}\left[\vX_t-R^-\,\vl_t/4\right]=0\label{forces-iso}\\
&&\vsigma\cdot\left[\vl_t+\delta
\mathbb{S}(R^-\vX_t+\kappa\vl_t)\right]=-2\varepsilon\alpha\vsigma\cdot\widetilde{\vg}/\kappa
\label{torq-iso}
\end{eqnarray}
where $R^-\equiv R_1-R_2$.

\section{Two-timing method and asymptotic procedure \label{sect04}}

\subsection{Functions and notations}

The following \emph{dimensionless} notations and definitions are in use:

\noindent
(i) $s$ and $\tau$ denote slow and fast times;  subscripts $\tau$ and $s$ stand for  related partial
derivatives.

\noindent
(ii) A dimensionless function, say $h=h(s,\tau)$, belongs to the class $\cal{I}$ if $h={O}(1)$ and all
partial $s$-, and $\tau$-derivatives of $h$ (required for our consideration) are also ${O}(1)$. In this paper
all functions belong to   class $\cal{I}$, while all small parameters appear as explicit multipliers.

\noindent
(iii) We consider only \emph{functions periodic in $\tau$} $\{h\in  \mathcal{P}:\quad
h(s,\tau)=h(s,\tau+2\pi)\}$, where $s$-dependence is not specified. Hence, all functions considered below
belong to $\cal{P}\bigcap\cal{I}$.

\noindent
(iv) For  arbitrary $h\in \cal{P}$ the \emph{averaging operation} is
\begin{eqnarray}
\langle {h}\,\rangle \equiv \frac{1}{2\pi}\int_{\tau_0}^{\tau_0+2\pi}
h(s, \tau)\,d\tau\equiv \overline{h}(s),\qquad\forall\ \tau_0\label{oper-1}
\end{eqnarray}

\noindent
(v) \emph{The oscillating part of an integral} is:
\begin{eqnarray}
&&\widetilde{h}^{\tau}\equiv\int_0^\tau \widetilde{h}(\vx,s,\nu)\,d\nu
-\frac{1}{2\pi}\int_0^{2\pi}\Bigl(\int_0^\mu
\widetilde{h}(\vx,s,\nu)\,d\nu\Bigr)\,d\mu\nonumber\label{oper-7}
\end{eqnarray}

\noindent
(vi)  \emph{The tilde-function} (or purely oscillating function) represents a special case of
$\cal{P}$-function with zero average $\langle\widetilde h \,\rangle =0$. The \emph{bar-function} (or
mean-function) $\overline{h}=\overline{h}(s)$ does not depend on $\tau$. For any periodic function $h$ a
unique decomposition $h=\overline{h}+\widetilde{h}$ is valid.

\subsection{Asymptotic procedure and successive approximations}

The introduction of a fast time variable $\tau$ and a slow time variable $s$ represents a crucial step in our
asymptotic procedure. We choose $\tau=t$ and $s=\varepsilon^2 t$. This choice can be justified by the same
distinguished limit arguments as in \cite{VladimirovMHD}. Here we present this choice without proof, however,
its most important part (that this choice  leads to a valid asymptotic procedure) is exposed and exploited
below. We use the chain rule
\begin{eqnarray}\label{chain}
&&d/dt=\partial/\partial\tau+\varepsilon^2\partial/\partial s
\end{eqnarray}
and then we  accept (temporarily) that $\tau$ and $s$ represent two independent variables. Further more we
consider series expansions in the small parameter $\varepsilon$ and restrict our attention to terms which are
at most $O(\varepsilon^2)$. Simultaneously, we keep at most linear in $\delta$ terms. It does not mean that
in our setting $\delta\sim\varepsilon^2$, since in all expressions $\delta$ appears not separately but as
products with various degrees of $\varepsilon$. Hence, we do not specify the dependence of unknown functions
on $\delta$; such dependence reveals itself naturally during the calculations. The unknown functions are
taken as regular series in $\varepsilon$
\begin{eqnarray}\label{x-f-ser}
&&\vX(\tau,s)=\vX_0(\tau,s,\delta)+\varepsilon \vX_1(\tau,s,\delta)+\varepsilon^2 \vX_2(\tau,s,\delta)+\dots
\end{eqnarray}
with a similar expression for $\varphi$.  We accept that
$$
\widetilde{\vX}_0(s,\tau,\delta)\equiv 0\  \text{and}\ \widetilde{\varphi}_0(s,\tau,\delta)\equiv 0
\quad\text{while}\quad \overline{\vX}_0(s,\delta)\neq 0\ \text{and}\ \overline{\theta}_0(s,\delta)\neq 0
$$
which express a basic property of our solutions that long distances of self-swimming and large angles of
rotation are caused by small oscillations.  The application of (\ref{chain}) to (\ref{x-f-ser}) gives
\begin{eqnarray}\label{x-f-ser1}
&&\vX_t=\varepsilon \widetilde{\vX}_{1\tau}+\varepsilon^2
(\widetilde{\vX}_{2\tau}+\overline{\vX}_{0s})+O(\varepsilon^2)
\end{eqnarray}
and a similar expression for $\varphi$. In the calculations below all bar-functions belong to the zero
approximation, while all  tilde-functions belong to the first approximation; therefore we omit the related
subscripts.

The successive approximations of equations (\ref{forces-iso}),(\ref{torq-iso}) yield:

\noindent \emph{Terms} $O(\varepsilon^0)$ give the identities $0=0$.

\noindent  \emph{Terms}  $O(\varepsilon^1\delta^0)$ lead to
\begin{eqnarray}
&&\widetilde{\vX}_{\tau}=0,\quad
\widetilde{\vl}_{\tau}\cdot\overline{\vsigma}=-2\alpha\widetilde{\vg}\cdot\overline{\vsigma}/\kappa
\label{first-order}
\end{eqnarray}
The use of (\ref{units}) transforms the second equation to the form
$\widetilde{\varphi}_{\tau}=-2\alpha\widetilde{\vg}\cdot\overline{\vsigma}/\kappa$. Then the integration of
(\ref{first-order}) in the class of periodic functions yields
\begin{eqnarray}
&&\widetilde{\vX}=0,\quad \widetilde{\varphi}=-2\alpha\widetilde{\vg}^\tau\cdot\overline{\vsigma}/\kappa
\label{first-order-2}
\end{eqnarray}

\noindent \emph{Terms} $O(\varepsilon\delta)$: These terms do not vanish and can be easily calculated.
However, as one can see below, they do not participate in the leading terms of the average motion that appear
in the orders $O(\varepsilon^2)$ and $O(\varepsilon^2\delta)$.

\noindent \emph{Terms} $O(\varepsilon^2)$: Eqns.(\ref{forces-iso})),(\ref{first-order-2}) give
\begin{eqnarray}\label{eqn-X-2}
&& \overline{\vX}_{s}=\frac{1}{2}\delta\kappa R^-\langle
\widetilde{\mathbb{S}}_\tau\widetilde{\vl}\rangle
\end{eqnarray}
where we have used the integration by parts in the average operation (\ref{oper-1}). The use of the
definition of matrix $\mathbb{S}$ (\ref{forces-Stokes}) and (\ref{units}) yield
$\langle\widetilde{\mathbb{S}}_\tau\widetilde{\vl}\rangle=
2\langle\widetilde{l}\widetilde{\varphi}_\tau\rangle\overline{\vsigma}$. Then (\ref{eqn-X-2}) takes the form
\begin{eqnarray}\label{eqn-X-4}
&& \overline{\vX}_{s}=-\delta\alpha R^-\langle
\widetilde{l}\widetilde{\vg}\cdot\overline{\vsigma}\rangle\overline{\vsigma}
\end{eqnarray}
Similarly,  from (\ref{torq-iso}),  we can derive the equation for $\overline{\varphi}$ and obtain the system
of equations
\begin{eqnarray}
&&\overline{\vX}_{s'}=-\mu{U}\overline{\vsigma}/\gamma,\quad \overline{\varphi}_{s'}={U}
-\gamma{G}; \label{X-phi-final}\\
&&s'\equiv \gamma s,\quad\mu\equiv \delta\alpha R^- ,\quad \gamma\equiv 2\alpha/\kappa,\quad{G}\equiv
\langle\widetilde{g}_1^\tau
\widetilde{g}_2\rangle\nonumber\\
&&{U}\equiv \langle
\widetilde{l}(\widetilde{\vg}\cdot\overline{\vsigma})\rangle=-{G_1}\sin\overline{\varphi}+{G_2}\cos\overline{\varphi},\quad
{G_1}\equiv\langle
\widetilde{l}\,\widetilde{g}_1\rangle,\quad
{G_2}\equiv\langle\widetilde{l}\,\widetilde{g}_2\rangle
\label{psi}
\end{eqnarray}
where  we have used the equality
$
\langle(\widetilde{\vg}^\tau\cdot\overline{\vn})(\widetilde{\vg}\cdot\overline{\vsigma})\rangle={G}
$, which is valid by virtue of (\ref{notations}),(\ref{oper-1}),  and
$\widetilde{\vg}=(\widetilde{g}_1,\widetilde{g}_2)$. One can see, that the dynamics of a dumbbell is
determined by the values of two parameters $\mu$, $\gamma$ and by three correlations ${G}$, ${G_1}$, and
${G_2}$. We can make a general conclusion, based on (\ref{X-phi-final}), that the mean translational velocity
is always perpendicular to the mean symmetry axes of a dumbbell ($\overline{\vX}_{s'}$ is directed along the
normal vector $\overline{\vsigma}$).

\section{Prescribed oscillations of the \emph{BD}-robot}

\subsection{Unidirectional oscillations of a fluid are not effective}

The simplest motion takes place when ${G}\equiv 0$. Physically, it means that fluid oscillations are
unidirectional, see (\ref{gravity}). In this case equations (\ref{X-phi-final}) produce the integral
\begin{eqnarray}\label{X-0}
\overline{\vX}-\vC=-\frac{\mu}{\gamma}\overline{\vn}\quad\text{or}\quad
(\overline{\vX}-\vC)^2=\mu^2/\gamma^2
\end{eqnarray}
with a vectorial constant of integration $\vC$. This equality shows that $\overline{\vX}(s')$ changes along a
circular path (or along an arc of a circle) of  small radius $\mu/\gamma=\delta \kappa R^-/2$. The equation
for $\overline{\varphi}$ (\ref{X-phi-final}) can be  integrated exactly. For  unidirectional oscillations
along the $y$-axis, when $\widetilde{\vg}=(0,\widetilde{g}_2)$ (\ref{gravity}), the second equation
(\ref{X-phi-final}) takes the form $\overline{\varphi}_{s'}={G_2}\cos\overline{\varphi}$. It can be
integrated, having  an initial value $\varphi(0)=\varphi_0$, as
\begin{eqnarray}
\sin\overline{\varphi}=\frac{(1+\sin\overline{\varphi}_0)e^{2G_2s'}-(1-\sin\overline{\varphi}_0)}
{(1+\sin\overline{\varphi}_0)e^{2G_2s'}+(1-\sin\overline{\varphi}_0)}
\end{eqnarray}
which shows that for $s\to\infty$ we have $|\overline{\varphi}|\to \pi/2$; it means that the axis of symmetry
of a dumbbell is turning monotonically towards the direction of oscillations. Equation (\ref{X-0}) describes
the simultaneous  change of $\overline{\vX}$ along the arc of a small circle. It is clear, that in the
general case of unidirectional oscillations along any  direction (different from $y$), the result is the
same: the axis of the dumbbell asymptotically approaches the direction parallel to the oscillations, and
$\overline{\vX}$ changes along the arc of a small circle. Therefore, we can conclude that any unidirectional
oscillations of the fluid do not result in the self-propulsion of the \emph{BD}-robot.

\subsection{Rectilinear self-propulsion without rotation}

For ${G}\neq 0$ we first consider motions without rotation $\overline{\varphi}_{s'}=0$. In these cases the
angular part of (\ref{X-phi-final}) gives
$-{G_1}\sin\overline{\varphi}+{G_2}\cos\overline{\varphi}=\gamma{G}$, which immediately leads to
\begin{eqnarray}
&&\overline{\varphi}={\varphi}_0=-\arctan({G_1}/{G_2})+\arccos(\gamma{G}/\sqrt{{G_1}^2+{G_2}^2})=\const\label{phi-line}\\
&&\text{when}\quad|\gamma{G}|\leq \sqrt{{G_1}^2+{G_2}^2}\label{phi-line-cond}
\end{eqnarray}
Physically, the restriction   (\ref{phi-line-cond})   means that the \emph{BD}-robot can move without
rotation if the oscillations $\widetilde{l}$ are `strong enough'. In this case the first eqn. in
(\ref{X-phi-final}) gives $\overline{\vX}_{s'}=-\mu{G}\overline{\vsigma}=\const$, which shows that the
\emph{BD}-robot moves with constant speed $|\mu{G}|$ in the fixed direction $\overline{\vsigma}$, which is given
by the angle ${\varphi}_0\pm \pi/2$ (\ref{phi-line}), where the sign is determined by the correlation ${G}$.
Striving for more general results, one can show, that the system (\ref{X-phi-final}) can be integrated
analytically in the general case $\overline{\varphi}_{s'}\neq 0$, with the conclusion, that if the parameters
satisfy (\ref{phi-line-cond}) then  a trajectory with any initial data $\overline{\varphi}(0)$ asymptotically
(when $s\to\infty$) approaches the same straight paths (\ref{phi-line}) as described above. Exact integration
outside of the range of parameters (\ref{phi-line-cond}) is also accessible analytically; it produces motions
with rotation $|\overline{\varphi}_s|>\const$, which we do not consider in this paper.

Let us consider a particular example
\begin{eqnarray}
&&\widetilde{g}_1=-a\sin\tau,\quad \widetilde{g}_2=b\cos\tau,\quad \widetilde{l}=l_0\sin\tau;
\quad\text{with}\quad \quad a>0,\ b>0,\
l_0>0\label{example}\\
&&{G_1}=-al_0/2,\quad {G_2}=0,\quad {G}=ab/2,\quad \varphi_0=\frac{\pi}{2}+\arccos(\gamma b/l_0)\nonumber
\end{eqnarray}
For $|\gamma b/l_0|\leq 1$ the \emph{BD}-robot propels itself with constant speed $|\overline{\vX}_{s'}|=\mu
a b/2$ along a straight path $\varphi=\varphi_0\pm\pi/2$.  It is remarkable, that the self-propulsion speed
$\mu a b/2$ does not depend on the amplitude $l_0$. However, one should keep in mind that such solutions are
available only for `strong' oscillations, when $|l_0|\geq |\gamma b|$; for `very strong' oscillations, when
$|l_0/b|\to\infty $, we have $\varphi_0\to\pi$.

\section{Elastic \emph{BD}-robot}

The above results correspond to the prescribed periodic function $\widetilde{l}(\tau)$ (\ref{constraint}),
which can be chosen arbitrary and represents a given time-dependent constraint. However, in practice, the
oscillations  $\widetilde{l}$ produced by the forces exerted from an oscillating fluid on the beads, are more
interesting. A simple way to consider such oscillations is to replace the rod with a spring of stiffness
$k=\const$. In this case the dimensionless Lagrangian function (\ref{Lagr-contstr}) and potential energy
(\ref{Pi}) become
\begin{eqnarray}
&&\mathcal{L}(\vq,\vq_t)=\mathcal{K}-\Pi,\quad \Pi=\varepsilon\alpha\widetilde{\vg}\cdot\vl+k(l-1)^2/2
\label{Lagr-contstr-A}
\end{eqnarray}
One can check that for this Lagrangian function the equations for total force and torque
(\ref{sum-forces}),(\ref{sum-torques}) remain the same, while  the equation for the reaction of constraint
(\ref{eqns-reactions}) must be replaced with
\begin{eqnarray}
&&\varepsilon\alpha\vg\cdot\vn+kl=-\kappa\vl_t\cdot\vn+O(\varepsilon\delta)
\label{eqns-reactions-A}
\end{eqnarray}
The relation between the problem for an elastic $BD$-robot and the previous problem for a $BD$-robot with the
arbitrary oscillation of a rod is evident: the latter considers all possible solutions, while the former
corresponds to a special subclass of $\widetilde{l}(\tau)$ only, that appears as the result of spring
oscillations. Hence the ability for self-propulsion can only `worsen' after the introduction of a spring. The
order $O(\varepsilon)$ equation (\ref{eqns-reactions-A}) produces a linear equation for $\widetilde{l}$
\begin{eqnarray}\label{eqn-l-A}
\widetilde{l}_\tau+K\widetilde{l}=-\gamma\widetilde{\vg}\cdot\overline{\vn},\quad K\equiv 2k/\kappa
\end{eqnarray}
It gives us $\widetilde{l}(\tau)$ which must be substituted into $G_1$ and $G_2$ in (\ref{X-phi-final})
instead of an arbitrary chosen function $\widetilde{l}$. The rest of the problem remains unchanged. The
general solution of (\ref{eqn-l-A}) can be obtained  analytically in an integral form, or in the form of a
Fourier series. Both forms are rather cumbersome and are not considered in this paper. Instead, we present an
example for a gravity field $\widetilde{g}_1=-a\sin\tau$, $\widetilde{g}_2=b\cos\tau$ that coincides with
(\ref{example}). The related solution of (\ref{eqn-l-A}) is
\begin{eqnarray}\nonumber
\widetilde{l}=\frac{1}{1+K^2}\left[(P+KN)\sin\tau+(KP-N)\cos\tau\right],\ P\equiv -\gamma b
\sin\overline{\varphi}, \
N\equiv \gamma a \cos\overline{\varphi}
\end{eqnarray}
where an exponentially decreasing complementary solution has been dropped. It leads to an explicit formula
for ${U}$ (\ref{psi})
\begin{eqnarray}\label{psi1}
{U}=\frac{\gamma}{2(1+K^2)}\left[ \frac{1}{2}K(a^2-b^2)\sin(2\overline{\varphi})-ab  \right]
\end{eqnarray}
which determines the system of equations (\ref{X-phi-final}). In this case we obtain the following equations
for the  motion without rotation ($\overline{\varphi}_{s}\equiv 0$)
\begin{eqnarray}\label{elastic-mot}
\overline{\vX}_{s'}=-\frac{1}{2}\mu a b\,\overline{\vsigma},\quad
\overline{\varphi}=\varphi_0=\frac{1}{2}\arcsin\Phi=\const,\quad \Phi\equiv\frac{2ab}{a^2-b^2} \left(\frac{2}{K}+K\right)
\end{eqnarray}
We can see, that any direction of a rectilinear self-propulsion can be arranged  by an appropriate choice of
$a$ and $b$. It is also interesting that the speed of self-propulsion, $\mu a b/2$, does not depend on the
spring stiffness $k$; however, the required condition $|\Phi|\leq 1$ shows that both a small and high
stiffness do not lead to rectilinear motion. Another interesting conclusion is: for rectilinear motion to
exist, the values of vibrational amplitudes $a$ and $b$ can not be chosen close to each other. It means that
imposed vibrations (\ref{gravity}) must be anisotropic (the circular vibrations with $a=b$ and close to them
are excluded). Again, for $\overline{\varphi}_{s'}\neq 0$, $|\Phi|\leq 1$ the system of equations
(\ref{X-phi-final}),(\ref{psi1}) can be integrated analytically. The integration shows that any trajectory
asymptotically (for $s\to\infty$) approaches (\ref{elastic-mot}). In the case $\Phi>1$ the full system
(\ref{X-phi-final}),(\ref{psi1}) also allows explicit analytical integration; it leads to the motions with
rotation $|\overline{\varphi}_s|>\const$, which we do not consider in this paper.

\section{Discussion}

(i)  Our choice of slow time $s=\varepsilon^2 t$ (\ref{chain}) agrees with classical studies of
self-propulsion for low Reynolds numbers, see \cite{Taylor, Blake, Childress}, as well as the geometric
studies of \cite{Wilczek}.

(ii) The scale of slow time $s=\varepsilon^2 t$ implies that in order to obtain physical dimensionless
velocities we have to multiply $\overline{\vX}_{s'}$ and $\overline{\varphi}_{s'}$ (\ref{X-phi-final}) by
$\gamma\varepsilon^2$. Accordingly, the  mean translational velocity $\overline{\vV}=O(\varepsilon^2\delta)$,
while mean angular velocity $\overline{\Omega}=O(\varepsilon^2)$.

(iii) We have built an asymptotic procedure with two small parameters: $\varepsilon\to 0$ and $\delta\to 0$.
Such a setting usually requires the consideration of different asymptotic paths on the plane
$(\varepsilon,\delta)$ when, say $\delta=\delta(\varepsilon)$. In our case we can avoid this additional
analysis, since $\delta$ does not appear separately,  but only in combinations like $\varepsilon^2\delta$.

(iv) In this paper we consider only plane motions of a three-dimensional dumbbell. This class of motions
corresponds to two-dimensional oscillations/gravity (\ref{gravity}).  At the same time, for experimental
realization, it could be necessary to solve a full three-dimensional problem.

(v) For the first experimental studies of self-propulsion of the \emph{BD}-robot one can consider: rigid-body
oscillations of a fluid, enclosed within a vibrating container; the viscous flows, caused by oscillatory
boundary conditions; or the oscillations of a fluid due to an external acoustic wave.

(vi) The calculated velocities of self-propulsion are much smaller than flow oscillations and can be even
smaller than some secondary flows (like acoustic streaming). In this case, one may look for situations when a
slow (but permanent) self-propulsion leading to new physical effects or providing  advantages in
applications.

(vii) It is well known, that an oscillating dumbbell is able to self-swim when an oscillating external
torque, exerted on a dumbbell, is present; the related discussion  can be found in
\cite{Felderhof, Felderhof1, Friedman}. The self-swimming of a magnetically driven oscillating dumbbell has been
studied by \cite{Gilbert}.

(viii) Our approach is technically different from all previous methods employed in the studies of
micro-robots.  The possibility to describe any motion of a \emph{BD}-robot explicitly, shows the strength and
analytical simplicity of our method. The studies of different micro-robots by the same method (as in this
paper) can be found in \cite{VladimirovX3, VladimirovX4, VladimirovX5}. In \cite{VladimirovX6} the same
method resulted in a new asymptotic model and a new equation (\emph{the acoustic-drift equation}) for the
averaged flows generated by acoustic waves.

\begin{acknowledgments}
The author is grateful to Profs. M.A. Bees, C.J. Fewster, P.H. Gaskell, A.D. Gilbert, K.I.  Ilin, H.K.
Moffatt, T.J. Pedley, and J.W. Pitchford for useful discussions.
\end{acknowledgments}


\begin{thebibliography}

\bibitem[Alexander \emph{et.al.}  (2009)]{Yeomans} \textsc{Alexander, G. P., Pooley, C. M., and Yeomans, J.M.}
2009 Hydrodynamics of linked sphere model swimmers. {\it J. Phys.: Condens. Matter}, \textbf{21}, 204108.

\bibitem[Alouges \emph{et.al.} (2008)]{Lefebvre} \textsc{Alouges, F., DeSimone, A., and Lefebvre, A.}
2008 Optimal strokes for low Reynolds number swimmers: an example. {\it J. Nonlinear Sci.},
\textbf{18}, 277-302.

\bibitem[Avron \emph{et.al.} (2005)]{Avron} \textsc{Avron, J.E., Kenneth, O., and Oaknin, D.H.}
2005 Pushmepullyou: an efficient micro-swimmer. {\it New J. of Physics},
\textbf{7}, 234.

\bibitem[Becker \emph{et.al.}  (2003)]{Koelher}
\textsc{Becker, D. J., Koelher, C. M., Ryder, and Stone, J.M.}
2003 On self-propulsion of micro-machimes al low Reynolds number: Purcell's three-link swimmer. {\it J. Fluid
Mech.}, \textbf{490}, 15-35.


\bibitem[Belovs \& C\"{e}rbers (2009)]{Belovs} \textsc{Belovs, M. and C\"{e}rbers, A.} 2009
Ferromagnetic microswimmer. {\it Phys.Rev.E}, \textbf{79}, 051503.


\bibitem[Blake  (1971)]{Blake}
\textsc{Blake, J. R.} 1971 Infinite models for ciliary propulsion. {\it J. Fluid Mech.},
\textbf{49}, 209-227.

\bibitem[Chang \emph{et.al.} (2007)]{Paunov}
\textsc{Chang, S.T., Paunov, V.N., Petsev, D.N., and Orlin, D.V.}
2007 Remotely powered self-propelling particles and micropumps based on miniature diodes. {\it Nature
Materials}, \textbf{6}, 235-240.

\bibitem[Childress (1981)]{Childress} \textsc{Childress, S.} 1981 \emph{Mechanics of swimming and flying.}
Cambridge, CUP.


\bibitem[Dreyfus \emph{et.al.} (2005)]{Dreyfus}
\textsc{Dreyfus , R., Baudry, J., Roper, M.L., Fermigier, M., Stone, H.A. and Bibette, J.}
2005 Microscopic artificial swimmers. {\it Nature},
\textbf{437}, 6, 862-865.

\bibitem[Earl \emph{et.al.} (2007)]{Yeomans1}
\textsc{Earl, D. J., Pooley, C. M., Ryder, J.F., Bredberg, I. and Yeomans, J.M.}
2007 Modelling microscopic swimmers at low Reynolds number. {\it J. Chem. Phys.},
\textbf{126}, 064703.

\bibitem[Felderhof (2006)]{Felderhof} \textsc{Felderhof, G.}
2006 The swimming of animalcules. {\it Physics of Fluids}, \textbf{18}, 063101.


\bibitem[Felderhof (2007)]{Felderhof1} \textsc{Felderhof, G.}
2007 Response to "Comment on  `The swimming of animalcules'". {\it Physics of Fluids}, \textbf{19}, 079102.

\bibitem[Friedman (2007)]{Friedman} \textsc{Friedman, B. U.}
2007 Comment on "The swimming of animalcules". {\it Physics of Fluids},
\textbf{19}, 079101.

\bibitem[Gilbert \emph{at.al.} (2010)]{Gilbert}
\textsc{Gilbert, A. D., Ogrin, F. Y., Petrov, P.G., and Wimlove, C.P.}
2010 Theory of ferromagnetic microswimmers. {\it Q.Jl Mech. Appl. Math.},
\textbf{64}, 3, 239-263.

\bibitem[Golestanian \& Ajdari (2008)]{Golestanian} \textsc{Golestanian, R. and Ajdari, A.} 2008
Analytic results for the three-sphere swimmer at low Reynolds number. {\it Phys.Rev.E}, \textbf{77}, 036308.




\bibitem[Landau \& Lifshitz (1959)]{Landau} \textsc{Landau, L.D. and Lifshitz, E.M.} 1959
\emph{Fluid Mechanics.} Oxford, Butterworth-Heinemann.

\bibitem[Lamb (1932)]{Lamb} \textsc{Lamb, H.} 1932 \emph{Hydrodynamics.} Sixth edition,
Cambridge, CUP.


\bibitem[Lauga (2011)]{Lauga}  \textsc{Lauga, E.} 2011 Life around the scallop theorem. {\it Soft Matter},
\textbf{7}, 3060-3065.



\bibitem[Leoni \emph{et.al.} (2009)]{Pietro}
\textsc{Leoni, M., Kotar, J., Bassetti, B., Cicuta, P. and Lagomarsino, M.C.}
2009 A basic swimmer at low Reynolds number. {\it Soft Matter},
\textbf{5}, 472-476.

\bibitem[Moffatt (1996)]{Moffatt} \textsc{Moffatt, H. K.} 1996 \emph{Dynamique des Fluides, Tome 1,
Microhydrodynamics.} Ecole Polytechnique, Palaiseau.

\bibitem[Najafi \& Golestanian (2004)]{NG+} \textsc{Najafi, A. and Golestanian, R.} 2004
Simple swimmer at low Reynolds number: three linked spheres. {\it Phys.Rev.E}, \textbf{69}, 062901.

%
%

\bibitem[Purcell (1977)]{Purcell} \textsc{Purcell, E.M.} 1977
Life at low Reynolds number. {\it Amer. J. of Phys.}, \textbf{45}, 1, 3-11.

\bibitem[Romanczuk \emph{et.al.}  (2012)]{Romanczuk}
\textsc{Romanczuk, P., B\"{a}r, M., Ebeling, W., Lindner, B, and Schimansky-Geier, L.}
2012 Active Brownian particles. {\it Eur. Phys. J. Special Topics}, \textbf{202}, 1-162.



\bibitem[Shapere \& Wilczek (1989)]{Wilczek} \textsc{Shapere, A. and Wilczek, F.} 1989
Efficiencies of self-propulsion at low Reynolds number. {\it J. Fluid Mech.},
\textbf{198}, 587-599.

\bibitem[Taylor (1951)]{Taylor} \textsc{Taylor, G.
I.} 1951, Analysis of the swimming of microscopic organisms. {\it Proc. R. Soc. Lond.}, \textbf{A209},
447-461.


\bibitem[Vladimirov (2005)]{Vladimirov0} \textsc{Vladimirov, V.A.}
2005  Vibrodynamics of pendulum and submerged solid.  {\it J. of Math. Fluid Mech.} \textbf{7}, S397-412.

\bibitem[Vladimirov (2008)]{Vladimirov1} \textsc{Vladimirov, V.A.}
2008 Viscous flows in a half-space caused by tangential vibrations on its boundary. {\it Studies in Appl.
Math.}, \textbf{121}, 4, 337-367.





\bibitem[Vladimirov (2012a)]{VladimirovMHD} \textsc{Vladimirov, V.A.}
2012a  Magnetohydrodynamic drift equations: from Langmuir circulations to magnetohydrodynamic dynamo?  {\it
J. Fluid Mech.} \textbf{698}, 51-61.

\bibitem[Vladimirov (2012b)]{VladimirovX3} \textsc{Vladimirov, V.A.} 2012b
Self-propulsion velocity of $N$-sphere micro-robot. Accepted to \emph{J. of Fluid Mech.}; E-print:
\emph{ArXiv}: 1206.0890v1 and 1209.0171v1, (physics,flu-dyn).

\bibitem[Vladimirov (2012c)]{VladimirovX4} \textsc{Vladimirov, V.A.} 2012c
Self-propulsion of V-shape micro-robot. E-print: \emph{ArXiv}: 1209.2835v1 (physics,flu-dyn).


\bibitem[Vladimirov (2012d)]{VladimirovX5} \textsc{Vladimirov, V.A.} 2012d Theory of a triangular micro-robot.
E-print: ArXiv: 1210.0747v1 (physics,flu-dyn).

\bibitem[Vladimirov (2012e)]{VladimirovX6} \textsc{Vladimirov, V.A.} 2012e Acoustic-drift equation.
E-print: ArXiv: 1206.1297v1 (physics,flu-dyn).


\end{thebibliography}
\end{document}